\documentclass{appolb}

\usepackage{graphicx}

\begin{document}
 \eqsec  % uncomment this line to get equations numbered by (sec.num)
\title{Experimental signals for broken axial symmetry  in excited heavy nuclei from the valley of stability
\thanks{Presented at XXIV Kazimierz NPW 20- 24 IX 2017\\ electronic address: E.Grosse@tu-dresden.de, a.junghans@hzdr.de }
}
\author{E. Grosse
%\altaffiliation{electronic address: E.Grosse@tu-dresden.de}
\address{Institute of Nuclear and Particle Physics, Technische Universit\"at Dresden, 01062 Dresden, Germany }
\\
{ A.R. Junghans}
%\altaffiliation{electronic address: a.junghans@hzdr.de}
 \address{ Institute of Radiation Physics, 	
Helmholtz-Zentrum Dresden-Rossendorf, 01314 Dresden, Germany}
}
\maketitle
\begin{abstract}
An increasing number of experimental data indicates the breaking of axial symmetry in many heavy nuclei already in the valley of stability:
Multiple Coulomb excitation analysed in a rotation invariant way, gamma transition rates and energies in odd nuclei, mass predictions, the splitting of Giant Resonances (GR), the collective enhancement of nuclear level densities and Maxwellian averaged neutron capture cross sections. For the interpretation of these experimental observations the axial symmetry breaking shows up in nearly all heavy nuclei as predicted by Hartree-Fock-Bogoliubov (HFB) calculations \cite{de10} ; this indicates a nuclear Jahn-Teller effect. We show that nearly no parameters remain free to be adjusted by separate fitting to level density or giant resonance data, if advance information on nuclear deformations, radii etc. are taken from such calculations with the force parameters already fixed. The data analysis and interpretation have to include the quantum mechanical requirement of zero point oscillations and the distinction between static vs. dynamic symmetry breaking has to be regarded. 

\end{abstract}
\PACS{PACS numbers come here}
  
\section{Introduction}
The enhancement seen in atomic hyperfine structure experiments for electric quadrupole (E2) moments \cite{sc35} over predictions for a configuration formed by only one or a few single particle orbitals has triggered intensive research on its consequences for nuclear structure: Besides the pressure exerted by the individual nucleons \cite{ra50} a theory of collective rotational motion as the origin of low energy $2^+$-states observed in even nuclei was derived \cite{bo53}. Also a successful spin predictions for low lying levels in odd nuclei became possible through the introduction of two oscillator frequencies for each main nuclear shell; this interplay between nucleonic motion and nuclear distortion was studied in nuclei away from magic shells possessing large equilibrium deformation. A possible breaking of axial symmetry in less deformed heavy nuclei was mentioned in these reviews, but the observation of additional $2^+$-levels was preferentially attributed to strongly excited dynamic shape changes along the symmetry axis ($\beta$ - vibration) or perpendicular to it ($\gamma$ - vibration) \cite{bo75}. An alternative explanation of the various quadrupole excitations by a rotation of a more static non-axial deformation was given \cite{df58}, but not persued by many other groups. The controversy between static vs. dynamic triaxiality is still discussed intensively, albeit the relation of nuclear deformation to the Jahn-Teller effect, well established as the cause of symmetry breaking in crystals \cite{ja37, re84}, may challenge the axial approximation as well as the spherical one. A possible reason for the often reported apparent nuclear axiality in Hartree Fock type variational calculations was shown to be related \cite{ha84} to the order for the projection to angular momentum in relation to the variational procedure (PAV vs.VAP) and hence questioned. In this paper we want to report on a survey in various fields of nuclear physics performed with the aim to test the possible influence of a breaking of axial symmetry on the analysis and interpretation of experimental results. We start with a short review on recent theoretical work and then we list experimental findings in heavy nuclei along the valley of stability which may give a hint on their symmetry.

\section{Axial symmetry in heavy nuclei}

\subsection{Recent theoretical work}
The importance of triaxiality in  nearly all heavy nuclei was also shown \cite{de10} by recent HFB-calculations using the Gogny D1S interaction, constrained to the selected values of Z and N and combined to the generator coordinate method. They use a triaxial oscillator basis with the product ${\omega_0}^3 ={\omega_x}{\omega_y}{\omega_z}$, where $\hbar\omega_0$ is obtained through minimization of the HFB energy. At this point $R_0$ and $R_p$ are defined as the charge and radius parameters for the equivalent mass resp. charge ellipsoid and the half axes are given by $R_i = \frac{\omega_0}{\omega_i\cdot R_p}$. Unfortunately different formulae have been proposed to convert half-axis values to deformation parameters which relate to observables. In a series of tests we determined that the axis lengths calculated from formulae in refs. \cite{hi53, ku72, de10} differ by less than 2 $\%$ for heavy nuclei when using identical deformations $\beta$ and $\gamma$; the relation to the convention based on spherical harmonics \cite{bo75} is more complicated as this is not volume-conserving. Following a suggestion \cite{an94}, we investigated a possible relation between the two deformation parameters for nuclei in the valley of stability \cite{gr17} and found a quite surprising correlation extending over the full range of deformation. The tabulated values from a recent Hartree-Fock-Bogolyubov calculation \cite{de10} were used; we repeat that this calculation is constrained to $A$ and $Z$ (CHFB) and is combined to the generator coordinate method (GCM) covering the full range of deformation. Assuming only $R_\pi$-invariance it predicts for many nuclei non-zero triaxiality $\langle\gamma\rangle \neq 0$, and in some cases the predicted standard deviation does not include $\gamma = 0$,  {\em i.e.} $\cos(3\gamma) = 1$. As was pointed out \cite{be07}, HFB-calculations tend to overpredict intrinsic electric quadrupole moments $Q_0$ for nuclei near closed shells, as they do not fully account for the very deep mean field potential there. Thus a reduction for nuclei only $\delta$ nucleons away from a shell a factor for the $\beta$-deformation \cite{de10} of $0.4 + \delta/20$ is applied for $\delta\leq 10$and this expression is used in our approach to describe GR shapes to be described later. We calculate correction factors for protons as well as neutrons and the larger of the two is taken; as an example we quote the resulting reduction of the predicted \cite{de10} $\beta$-values by 40, 30, 20 and $10 \%$ for the isotopes $^{148-154}$Sm. We thus use tabulated deformation values \cite{de10} for $cos 3 \gamma $ and the corrected $\beta$ to obtain $Q_0$; in Fig.1 they are plotted against each other. 

%uncomment the following lines to place a figure
\begin{figure}[htb]
\centerline{%
\includegraphics[width=8cm]{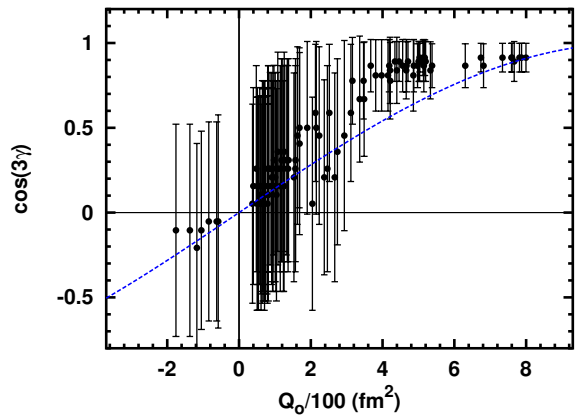}}
\caption{Plot of $cos 3 \gamma $ vs. $Q_0$ as obtained from CHFB calculations \cite{de10}; the data as well as the eye-guide (blue dash) indicate a rise in the axiality with increasing quadrupole deformation.}
\label{Fig.1}
\end{figure}

A rather clear correlation is obvious \cite{an94}, but the apparent tendency versus triaxiality for nuclei with small $Q_0$ (like the isotopes of Pb) may be considered surprising. It becomes less confusing when the rotation invariant $Q_0^3   cos 3\gamma$ is used as ordinate representing the dependence of axiality versus the quadrupole moment; it would tend more strongly towards zero for non-deformed nuclei. Rotation invariants were originally introduced \cite{ku72} for $^{152}$Ba, a nucleus with intermediate deformation, but they are of general value: The relation of the observable B(E2) to the deformation $\beta$ may vary with increasing spin \cite{ri82}, in contrast to such invariants. This is very helpful especially in the analysis of experimental multiple Coulomb excitation data \cite{cl86, sr11}.

\subsection{Ground state masses of heavy nuclei}
Whereas nuclear theory studies other than the one mentioned above \cite{de10} often still prefer to assume axial symmetry {\em  ad hoc} we actually question that an experimental proof for such symmetry exists, and probably all heavy nuclei do not strictly obey it, in apparent resemblance to what was found for tetrahedral crystals by Jahn and Teller in 1937 \cite{ja37}, only much later observed and demonstrated to also apply to sphericity and axiality of nuclei \cite{re84}. The suggested analogy is very reasonable as nuclei are three-dimensional objects like crystals and hence the axial description in two dimensions must be an approximation which has to be justified whenever used. A global test was performed by the extended Thomas-Fermi plus Strutinsky integral (ETFSI) method generalized to include a possible triaxiality of the nuclear shape \cite{du00}. In this paper ground state masses are calculated for 36 randomly selected nuclides from the valley of stability and tested with respect to a lowering as compared to an axial approximation; such a lowering was found for all of  them. Albeit this effect turns out to be rather small the results are at variance to various predictions as reviewed \cite{mo08}; the latter also presents a Table presenting the results obtained by the Finite-Range Liquid-Drop Model (FRLDM). Unfortunately this table only lists   nuclei for which the calculated decrease in energy due to triaxiality is equal to or larger than 0.01 MeV and this is misleading as it may be misunderstood as a proof of the axiality of the others. The sensitivity of the ground state mass calculated in the FRLDM against triaxiality is weak, but axial symmetry may well be broken even if its effect is below the calculational accuracy.

\subsection{Level energies and transition rates}
In a recent study \cite{na17} deviations between FRLDM and $\gamma$-values from gamma-decay spectra for $^{116-118}$Ru - obtained from fission product spectrocopy - were pointed out: The measured “signature” splitting of the yrast bands, when compared with the Triaxial Projected Shell Model (TPSM) calculations, shows the need for large, nearly constant, triaxial deformations near 30 degrees, which differ considerably from the FRLD-predicitions. Similar discrepancies to theoretical values for were reported \cite{me74} already long ago: In three odd nuclei close to $^{208}$Pb the observed level energies agree well to a calculation based on a single-j nucleon coupled to an asymmetric rotor with $\gamma$ clearly diffferent from an axial rotor prediction. The authors suggest that triaxial minima probably are more pronounced in heavy nuclei than predicted by existing calculations. Later on various groups investigated the possibity of triaxialiy in various nuclides by extracting information on $\gamma$ from spectroscopic observables: In one case \cite{wu96} it was determined for 25 deformed heavy nuclei with $150 <A <238$ from transition energies as well as from B(E2)-values. As reasonable agreement between the two was found, they state to have demonstrated quantitatively that the rms shape of all these nuclei is triaxial. 

Another group \cite{zh99} has pursued similar ideas covering a "broad range of nuclei from Z = 50-82, including nondeformed, deformed, and $\gamma$-soft nuclei". They first use results from an analysis of experimentcal data based on the Davydov model \cite{df58} which describes the rotation of a triaxial rigid body: $\gamma_E$ is taken from energy ratios and $\gamma_{(BE2)}$ as well as $\gamma_{(br)}$ are derived from the B(E2)-values or ratios, taken from three low lying 2$^+$ states. These are compared against each other and in a second step also with $\gamma_Q$ obtained from fits to experimental data using the interacting boson approximation (IBA). The authors find that the agreement or correlation in the pairwise comparisons of $\gamma$-values is reasonable; they also state that a rotation-invariant approach provides an approximate validation of the extraction of empirical triaxiality-values from the Davydov model. But they  apparently wonder - in contrast to the work mentioned before - if its inherent rigidity concerning axial asymmetry creates a bias versus static triaxiality in such a study. 

Recently work was published \cite{be10} where ground-state deformation parameters $\beta$ and $\gamma$ for stable Kr, Xe, Ba, and Sm isotopes were calculated using the eigenvalues of corresponding IBA-1 calculations, and the resulting modifications of the equilibrium deformations as taken from LDM compilations \cite{mo08} were listed. With the exception of two well deformed nuclei $\gamma$-values between 17 and 30 deg  found and this may be considered an indication of triaxiality, be it static or dynamic. This we consider a clear indication to look for other observables of relevant sensitivity, and this is expected to be best for not strongly deformed nuclei.were

\subsection{Coulomb excitation and reorientation}
The BE2-values and quadrupole moments of nuclei with given intrinsic deformation can be derived microscopically as solutions of the cranked HFB-equations \cite{ri82}. They are similar to those calculated for a rigid triaxial rotor \cite{df58} with the same deformation; both observables depend differently on the intrinsic axial asymmetry. Thus it is an evident strategy to measure both observables in one nucleus with sufficient accuracy to get direct information on its possibly broken axial symmetry. Unfortunately the experimental values reported for Q($2^+$) have rather large uncertainties related to difficulties in the measurement of Coulomb excitation reorientation. But in the case of $^{204}$Pb and $^{206}$Pb the B(E2)-values and hence their influence on the determination of $\gamma$  are very small, such that the respective paper \cite{jo78} gives values of $\gamma=43(8)^{\circ}$ and $ 33(6)^{\circ}$ clearly indicating static triaxiality with a tendency versus oblateness. This is a very interesting result with respect to the very often made assumption of near magic nuclei being rigidly spherical.

In the combination of measurements for Coulomb excitation yields obtained with projectiles with different Z and their correlated analysis respecting the effect of rotation invariants the extraction of the deformation parameters $\beta$ and $\gamma$ becomes less ambiguous. This was demonstrated \cite{cl86, wu95} for a large number of heavy nuclei, and the analysis even allowed to estimate the equantum mechanical zero-point oscillation of the values derived. Such investigation of nuclei with $180<A<200$ "clearly elucidates the smooth transition from prolate strongly-deformed shapes to less deformed triaxial shapes that have considerable softness to triaxial vibrations" as stated by authors. Even for 25 rather well deformed nuclei intrinsic E2 matrix elements  have been deduced from measured interband E2 matrix elements between ground and $\gamma$ band \cite{wu96, sr11}.  After correcting for the angular momentum dependence of the coupling between the rotation and intrinsic motion centroids for the possibly fluctuating triaxiality are obtained, and these correlate well with the values obtained from the analysis of excitation energies, based on rigid triaxiality \cite{df58}.
  
\subsection{Collective enhancement of level densities}
The statistical model of nuclear reactions derives the exit channel phase space from the  density of levels in the produced nuclei. A first estimate \cite{be37} was derived from the assumption that nuclear excitation can be modelled like a Fermi gas. The importance of rotational modes at low energy - well known from spectroscopy - lead to modifications \cite{er58, gi65}. A generalisation to non-axial deformation is clearly indicated in view of the various band structures found in nuclear spectroscopy experiments; it was proposed to include it in a group theoretical approach \cite{bj74} which handles the possible symmetries and their breaking. The Fermi gas state density, valid in the body fixed reference frame, is well defined, when the level density parameter $\tilde{a}$ from nuclear matter and a critical temperature $t_{ph-tr}$ from Fermi gas theory are applied. 

The parity independent level density, observable in the laboratory, is obtained for every spin I of an even nucleus by multiplication with a factor depending on the shape symmetry, which is $(2I+1)/4$, if nothing but the conservation in a rotation by $\pi$ is assumed. It is larger if axial deformation is assumed and even more so for the triaxial case \cite{bj74}. By a spin cut-off term a correction for the rotational energy is assured, and only shell and pairing effects have to be known for a prediction of level densities, when the shape symmetry is known . A sensitive test of this prescription is possible just above the neutron capture energy $S_n$ from the capture resonance spacings observed by neutron time of flight. For spin 0 target nuclei it was carried out by us recently \cite{gr14, gr17} and the results are depicted in Fig. 2, which also shows the reduction predicted for an assumption  of axial or spherical symmetry. The discrepancies for $A>140$ may be related to octupolar deviations from $R_\pi$-symmetry.

\begin{figure}[htb]
\centerline{%
\includegraphics[width=8cm]{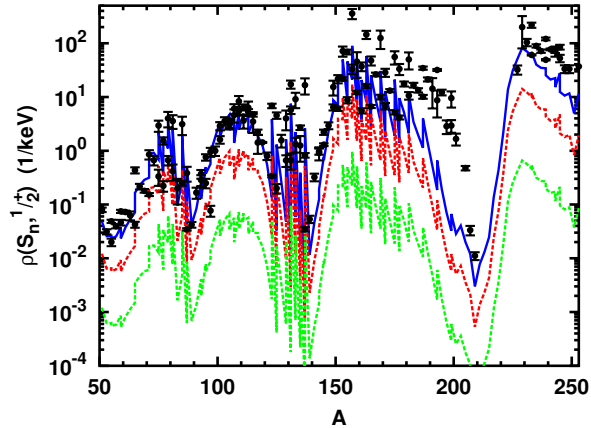}}
\caption{Plot of the level density near $S_n$ \cite{ca74} vs. A for even-odd nuclei in the valley of stability shown as black bars. The drawn blue curve shows the result of our parameter-free predicition assuming only the symmetry related to $R_\pi$, a rotation by $180^{\circ}$. The dashed (red) curve visualizes the effect of the axial symmetry approximation and the lowest (dashed green) curve was calculated without any collective enhancement \cite{be37} and Thomas-Fermi spin dispersion \cite{je52}.}
\label{Fig.2}
\end{figure}

The strong change with A was found to be related to the shell correction reducing the ground state mass as compared to LDM or Thomas-Fermi predictions; effects of nucleon pairing play an increasing role with decreasing excitation energy $E_x$ and lead to a steepening in the slope of level density vs. $E_x$ below the transition to a quasi superfluid nuclear phase. As we obtain the absolute normalization from the assumption that a Fermi gas prescription based on a nuclear matter Fermi energy is valid down to the phase transition point, we get a parameter free formula and we do not consider $\tilde{a}$ as a freely adjustable "level density parameter". Some ambiguity remains in the actinide region where 1. the shell correction is difficult to fix and 2. the symmetry about $\pi$ might be broken. But aside from that we can state that the breaking of axial symmetry is clearly indicated by the experimental data albeit they do not allow a determination of the absolute value of $\gamma$.

\subsection{Splitting of Giant resonances}
The splitting of the Isovector Giant Dipole Resonance (IVGDR) is proposed as an indicator of axial deformation in many nuclear physics textbooks like in the one of A. Bohr and B. Mottelson \cite{bo75}; there a local fit to the experimental data for the five isotopes regarded was performed and this indicated an independent adjustment of the apparent width to be obviously superior. In the work of our group recently reviewed \cite{ju17, gr18} a more rigorous approach was pursued by applying the old idea of the width to be mainly given by the resonance's energy without a local adjustment for each isotope eventually dependent of $A$ and $Z$. Here the modification accounting for triaxiality \cite{bu91} was helpful to obtain a consistent fit for 23 nuclei (as examples) in a wide range of mass number A under the assumption of broken axial symmetry $\Gamma_i = c_w E_i^{1.6}$. This triple Lorentzian procedure \cite{ju10,er10,gr18} (TLO) enables a restriction to only one global fit parameter $c_w = 0.045(3)$, which is valid for all nuclei regarded, to parametrize the IVGDR-width; actually this energy dependence is also valid to describe the width variation in one nucleus with its three pole energies taken from the three oscillator frequencies predicted \cite{de10} by the HFB calculation.

Actually our respective prediction for ${}^{208}$Pb agrees to the predicted \cite{do72} spreading width and we could disregard our fitting for $c_w$; the energy integrated absorption cross section is also fixed, as we require agreement to the TRK sum rule. The central resonance energy $E_0$ is fixed via the LDM \cite{my77} and an adjusted effective nucleon mass $m_{\rm eff} c^2$=800 MeV; other fits to the data only involve the peak widths, which will now be discussed. In Fig. 3 results for two neighboring nuclei are shown to visualize that for ${}^{150}$Sm we clearly need three poles when we use the same width parameter as in $^{152}$Sm; only for the latter a reasonable fit is also possible when using the axial approximation \cite{ca09}, as is well known \cite{df58} for strongly deformed nuclei. A modification resulting from sampling the IVGDR shapes according to the variances given \cite{de10} for the CHFB-calculation does not influence that conclusion.

\begin{figure}[htb]
\centerline{%
\includegraphics[width=12.5cm]{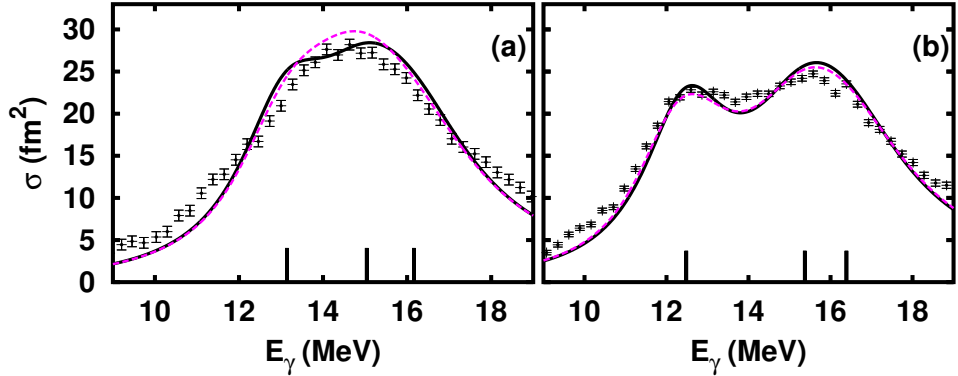}}
\caption{Plot of the photoneutron cross section \cite{ca74} for $^{150}$Sm (a) and $^{152}$Sm (b) together with the TLO sum of three Lorentzians (drawn curve) with $E_i$ indicated as black bars. The dashed (purple) curves visualizes the effect of shape sampling \cite{zh09, er10, gr17}.} 
\label{Fig.3}
\end{figure}

\subsection{Photon strength and neutron capture}
The radiative capture of fast neutrons by heavier nuclei plays an important role in considerations for advanced nuclear systems and it is of interest also for the cosmic nucleosynthesis. To test the combination of the present ansatz on the photon strength to the one for the level density - both allowing for a breaking of axial symmetry - a comparison on absolute scale of predicted to measured average radiative widths is possible. A sum over the decay channels to all bound states $J_b$ which can be reached from the capture resonances $J_r$ by photons of energy $E_\gamma = E_r - E_b$, multiplied by their density $\rho(E_b,J_b)$ leads to an effective averaging and to a maximum sensitivity of the product to rather low photon energy. 

As shown by us earlier \cite{sc12, gr17} the impact of photon strength on radiative neutron capture cross sections is peaking in the region below $E_\gamma \cong 5$ MeV and low energy modes may have some effect as well as irregular $A$ and $Z$ dependence of $\Gamma_\gamma$ or modifications in its slope vs. $E_\gamma$ \cite{ca09}. In our TLO approach the sole variation of $\Gamma_i$ with the pole energies $E_i$ avoids such problems by a strict implementation of the TRK sum rule: At least within the valley of stability a good agreement is obtained on absolute scale for neutron capture in the range of unresolved resonances. This is depicted in Fig. 4 for the Maxwellian average cross sections compiled recently \cite{di10} for $\langle E_n \rangle = 30 $~keV. An essential feature here is our global ansatz for the spreading width which fixes the important tail of the E1-resonance; it depends on the IVGDR pole energies only and their dependence on exact deformation parameters is nearly unimportant. But the broken axial symmetry has a large influence on the absolute value of the density of levels reached in the capture process. Thus the good agreement to experimental data as seen in the figure can be considered a clear support of our preference for non-axiality. 

\begin{figure}[htb]
\centerline{ 
\includegraphics[width=8cm]{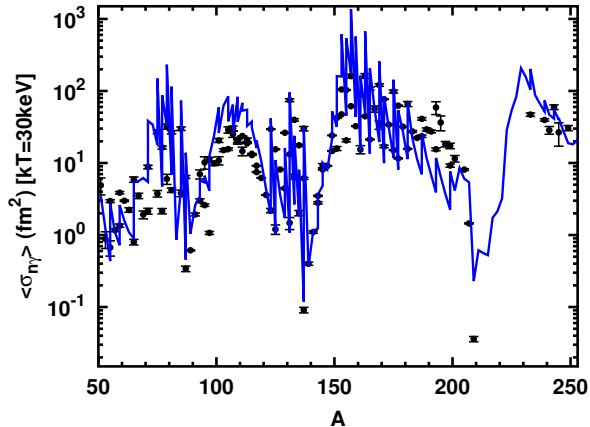}}
\caption{Plot of the Maxwellian averaged neutron capture cross section shown as black dots \cite{di10} for even-even nuclei together with the TLO ansatz using three Lorentzians combined to the discussed parameter free level density prediction (drawn curve) versus $A$ in the valley of stability. }
%The dashed (purple) curves visualizes the effect of an extra rise proposed earlier \cite{ka83, ko90}.} 
\label{Fig.4}
\end{figure}

For very neutron rich nuclei with their small $S_n$ the situation may become more complex, and experimental tests may become possible eventually in newly available radioactive beam facilities.

\section{Conclusion}
Various spectroscopic information presented over the years \cite{me74, ri82, cl86, an94, wu96, na17} indicated triaxiality for a number of heavy nuclei. Admission of the breaking of axial symmetry in accord to CHFB calculations \cite{de10} clearly improves a global description of Giant Dipole Resonance (IVGDR) shapes by a triple Lorentzian (TLO), introduced recently \cite{ju08, ju10, gr11} and discussed in some detail by this paper. The three parts add up to the TRK sum rule, when theoretical predictions for the $A$-dependence of pole energies from droplet model \cite{my77} and spreading widths based on one-body dissipation \cite{bu91} are used. The consideration of broken spherical and axial symmetry – for low excitation as well as for increasing excitation energy –, even when only weak, allows for a surprising reduction of the number of free fit parameters in two fields: The photon strength function and also in our novel approach to the density of low spin states populated by neutron capture. 

Thus a combination of the ansatz for photon strength and the one for level densities leads to a surprisingly good prediction of  Maxwellian averaged capture cross sections for more than $100$ spin-0 target nuclei with $ A>50$. They, as well as resonance spacing data are well described by a global ansatz with all parameters adjusted in advance and independent of the respective quantities. The triple Lorentzian (TLO) fit to IVGDR`s is global as it has only one free parameters – an effective nucleon mass – adjusted simultaneously to many resonance energies. The width can be taken from a HFB calculation \cite{do72} for $^{208}$Pb and adjusted to other $A$ and $Z$ only indirectly via $E_i$; the strength integrated over the IVGDR follows the TRK sum rule, which hence is always fulfilled. We stress that our good representation of level densities and IVGDR's together with the previous multiple Coulomb excitation data from the Rochester-Warsaw collaboration clearly hint versus a rigidity of triaxiality; especially for nuclei with intermediate deformation. Other data like the chiral effects in odd nuclei as well as the collective enhancement of level densities had already proven to be at variance with the often made assumption of axial deformation in heavy nuclei.

%%%%% CLEAR DOUBLE PAGE!
\newpage{\pagestyle{empty}\cleardoublepage}

\begin{thebibliography}{100}

	\bibitem{de10} J.-P. Delaroche et al., Phys. Rev. C 81, 014303 (2010), suppl. material

	\bibitem{sc35} H. Schiller and Th. Schmidt, Zeits. f. Phys. 94 (1935) 457

	\bibitem{ra50} J. Rainwater, Phys. Rev. 79, 432 (1950)

	\bibitem{bo53} A. Bohr and B.R. Mottelson, Mat. Fys. Medd. Dan. Vid. Selsk. 27, no. 16 (1953)

	\bibitem{bo75} A. Bohr and B. Mottelson, Nuclear Structure ch. 6, (New York 1975)

	\bibitem{df58} A.S. Davydov and J.P. Fillipov, Nucl. Phys. A 8, 237 (1958)

	\bibitem{ja37} H.A. Jahn and E. Teller, Proc. Roy. Soc. A161, 220 (1937)

	\bibitem{re84} P.G. Reinhard and E.W. Otten , Nucl. Phys. A 420, 173 (1984)

	\bibitem{ha84} A. Hayashi, K. Hara and P. Ring, Phys. Rev. Lett. 53, 337 (1984)

	\bibitem{hi53} D.L. Hill and J.A.Wheeler, Phys. Rev.  89, 1102 (1953)

	\bibitem{ku72} K. Kumar, Phys. Rev. Lett. 28, 249 (1972)

	\bibitem{an94} W. Andrejtscheff and P. Petkov, Phys. Lett. B 329, 1 (1994)

	\bibitem{gr17} E. Grosse, A.R. Junghans and J. Wilson, Phys. Scr. 92, 114003 (2017)

	\bibitem{be07} G.F. Bertsch et al., Phys. Rev. Lett. 99, 032502 (2007)

	\bibitem{ri82} P. Ring, A. Hayashi, K. Hara, H. Emling and E. Grosse, Phys. Lett. B 110, 423 (1982)

	\bibitem{cl86} D. Cline, Ann. Rev. Nucl. Part. Sci. 36, 683 (1986)

	\bibitem{sr11} J. Srebnry et al., Int. J. of Mod. Phys. E 20, 422 (2011)

	\bibitem{du00} A.K. Dutta, J.M. Pearson, and F. Tondeur, Phys. Rev. C 61, 054303 (2000)

	\bibitem{mo08} P. M\"{o}ller et. al., At. Data and Nucl. Data Tables  94, 758 (2008)

	\bibitem{na17} A. Navin et al., Phys. Lett. B 767, 480 (2017)

	\bibitem{me74} J. Meyer-ter-Vehn, et al., Phys. Rev. Lett. 32, 383 (1974)

	\bibitem{wu96} C. Y. Wu  and D. Cline, Phys. Rev. C 54, 2356 (1996)

	\bibitem{zh99} J. Zhang et al., Phys. Rev. C 60, 021304 (1999)
	
	\bibitem{zh09} S.Q. Zhang et al., Phys. Rev. C 80, 021307 (2009)
  
	\bibitem{be10} I. Bentley et al., Phys. Rev. C 83, 014317 (2011)

	\bibitem{jo78} A.M.R. Joye et al, Phys. Lett. B 72, 307 (1978)

	\bibitem{wu95} C. Y. Wu  and D. Cline, Nucl. Phys. A 607, 178 (1996)

	\bibitem{be37} H.A. Bethe, Phys. Rev. 50, 332) 1937)

	\bibitem{er58} T. Ericson, Nucl. Phys. 6, 62 (1958)

	\bibitem{gi65} A. Gilbert and A. G. W. Cameron, Can. Journ. of Phys. 43, 1446 (1965)

	\bibitem{bj74} S. Bj\o{}rnholm et al.,, Rochester-conf., IAEA-SM-l74, 205 (1974)

	\bibitem{gr14} E. Grosse, A.R. Junghans and R. Massarczyk, Phys. Lett. B 739, 1 (2014)

	\bibitem{ca74} P. Carlos et al., Nucl. Phys. A 225,171 (1974)

	\bibitem{je52} J.H.D. Jensen and J.M. Luttinger, Phys. Rev. 86, 907 (1952)

	\bibitem{ju17} A.R. Junghans et al., Eur. Phys. J. A, Web of Conf. 146, 05007 (2017)

	\bibitem{gr18} E. Grosse, A.R. Junghans and R. Massarczyk, Eur Phys. J. A, accepted

	\bibitem{bu91} B. Bush and Y. Alhassid, Nucl. Phys. A 531, 27 (1991)

	\bibitem{ju10} A.R. Junghans et al., Journ. Korean Phys. Soc. 59, 1872 (2010)

	\bibitem{er10} M. Erhard et al, Phys. Rev. C 81, 034319 (2010)

	\bibitem{do72} C.B. Dover et al., Ann. Phys. (N.Y.) 70, 458 (1972).

	\bibitem{my77} W.D. Myers et al., Phys. Rev. C 15, 2032 (1977)

	\bibitem{ca09} R. Capote et al., Nucl. Data Sheets 110, 3107 (2009)

	\bibitem{sc12} G. Schramm et al., Phys. Rev. C 85, 014311 (2012)

	\bibitem{di10} I. Dillmann et al., Phys. Rev. C 81, 015801 (2010), http://www.kadonis.org

	\bibitem{ju08} A.R. Junghans et al., Phys. Lett. B 670 (2008) 200

	\bibitem{gr11} E. Grosse et al., Eur. Ph. Journ. Web of Conf. 21, 04003  (2012)


\end{thebibliography}
\end{document}